\documentclass[aps,amsmath,twocolumn,amssymb,floatfix,superscriptaddress,footinbib]{revtex4-1}
\pdfoutput=1
\usepackage[dvips]{graphics}
\usepackage{bm}
\usepackage{epsfig}
\usepackage{enumerate}
\usepackage{subfigure}
\usepackage{color}
\usepackage{amsmath,amssymb,amsthm}
\usepackage{sidecap}
\usepackage{graphicx}
\usepackage{makecell}
\usepackage{hyperref}
\usepackage{multirow}
\usepackage{soul}

\usepackage[normalem]{ulem}

\hypersetup{
    pdfnewwindow=true,      
    colorlinks=true,       
    linkcolor=blue,          
    citecolor=magenta,        
    filecolor=blue,      
    urlcolor=blue        
}

\newcommand{\bea}{\begin{eqnarray}}
\newcommand{\eea}{\end{eqnarray}}
\newcommand{\beq}{\begin{equation}}  
\newcommand{\eeq}{\end{equation}}

\begin{document} 
\title{Dia- and paramagnetic Meissner effect from odd-frequency pairing in multi-orbital superconductors}
\author {Fariborz Parhizgar}
\affiliation{Department of Physics and Astronomy, Uppsala University, Box 516, S-751 20 Uppsala, Sweden}
\author {Annica M. Black-Schaffer}
\affiliation{Department of Physics and Astronomy, Uppsala University, Box 516, S-751 20 Uppsala, Sweden}

\begin{abstract}
The Meissner effect is one of the defining properties of superconductivity, with a conventional superconductor completely repelling an external magnetic field. In contrast to this diamagnetic behavior, odd-frequency superconducting pairing has often been seen to produce a paramagnetic Meissner effect, which instead makes the superconductor unstable due to the attraction of magnetic field.
In this work we study how both even- and odd-frequency superconducting pairing contributes to the Meissner effect in a generic two-orbital superconductor with a tunable odd-frequency pairing component. By dividing the contributions to the Meissner effect into intra- and inter-band processes, we find that the odd-frequency pairing actually generates both dia- and paramagnetic Meissner responses, determined by the normal-state band structure. More specifically, for materials with two electron-like (hole-like) low-energy bands we find that the odd-frequency inter-band contribution is paramagnetic but nearly canceled by a diamagnetic odd-frequency intra-band contribution. Combined with a  diamagnetic even-frequency contribution, such superconductors thus always display a large diamagnetic Meissner response to an external magnetic field, even in the presence of large odd-frequency pairing.
For materials with an inverted, or topological, band structure, we find the odd-frequency inter-band contribution to instead be diamagnetic and even the dominating contribution to the Meissner effect in the near-metallic regime. Taken together, our results show that odd-frequency pairing in multi-orbital superconductors does not generate a destabilizing paramagnetic Meissner effect and can even generate a diamagnetic response in topological materials.
\end{abstract}

\maketitle

\section{Introduction}\label{sec:Intro}

Odd-frequency superconductivity is an unusual superconducting state where the two electrons forming a Cooper pair join each other at different times, such that the resulting pair amplitude is odd under the interchange of the internal time coordinate, or equivalently, has an odd frequency dependence \cite{Berezinskii1974, Linder2019Rev, cayao2020Oned, Bergeret2005Rev}. 
The oddness under time interchange has direct consequences for the other symmetry properties of the superconducting state, since the fermionic nature of the pair amplitude requires it to be odd under a full interchange of all the electron quantum numbers. For example, the conventional superconducting spin-singlet $s$-wave state can be converted into a spin-triplet $s$-wave state if the frequency dependence becomes odd. This particular situation has been shown to arise when a conventional superconductor is in proximity with a ferromagnet, where the spin-triplet pairing allows for a long-range proximity effect into the ferromagnet \cite{Bergeret2005Rev, BergeretPRL2001,Volkov2003}.

More recently, odd-frequency superconductivity was also shown to occur in materials with an additional low-energy degree of freedom \cite{Triola2020reviewAnnPhy}, such as multi-orbital or band \cite{Annica2013multiband,Johann2020}, double dot \cite{Burset2016, Sothmann2014}, or bilayer structures \cite{Parhizgar2014}. In these systems the orbital (or similar) degree of freedom adds an additional index to the pair amplitude, such that it is, e.g., possible to have odd-frequency pairing in the form of spin-singlet $s$-wave odd inter-orbital pairing. Notably, this odd-frequency pairing easily occurs in the bulk and does not require any spatial inhomogeneity in form of junctions or similar. The condition to generate odd-frequency pairing in a multi-orbital superconductor \cite{Annica2013multiband, Triola2020reviewAnnPhy} is in fact easily fulfilled as it only requires finite inter-orbital (single particle) hybridization and that the strength of the superconducting pairing is different in different orbitals, also known as a pairing asymmetry or orbital selectivity \cite{Nica2017orbital}. As such, many known multi-band superconductors have recently been shown to host odd-frequency pairing, including doped topological insulators \cite{Parhizgar2017scirep, Johann2020}, Sr$_2$RuO$_4$ \cite{Lucia2017Kerr}, UPt$_3$ \cite{Chris2018Kerr}, and superconducting Weyl semimetals \cite{Parhizgar2020Jos, Paramita2020Jos}. Based on these results, even the iron-based superconductors with orbital selective pairing \cite{Nica2017orbital} likely host odd-frequency pairing.

As odd-frequency pairing is intrinsically time dependent it has so far been hard to detect directly, becoming essentially a hidden dynamic order. However, for specific systems signatures of odd-frequency pairing have been found in experimentally accessible quantities, such as the existence of finite density of states at zero energy \cite{Tanaka2007}, Kerr effect \cite{Lucia2017Kerr, Chris2018Kerr}, or Josephson current in junctions where even-frequency superconductivity becomes forbidden \cite{Eschrig2008triplet, Asano2017Jos, Parhizgar2020Jos, Paramita2019Weyl}. Most prominent of the odd-frequency signatures is  probably however the prediction of a paramagnetic Meissner effect \cite{Abrahams1995, Yokoyama2011, Alidoust2014triplet, Fominov2015, Bernardo2015Meis, Hoshino2014Meis, Lee2017Meis, Krieger2020Meis}.

The Meissner effect is the response of a superconductor to a weak magnetic field, with a conventional superconductor always completely repelling the magnetic field, known as a diamagnetic Meissner effect \cite{Meissner1933}.
In contrast to this conventional diamagnetic response, odd-frequency pairing has instead in many circumstances been predicted to cause a paramagnetic Meissner effect, where the superconductor instead attracts a magnetic field, which then also easily destabilizes the whole superconducting state \cite{Abrahams1995, Yokoyama2011, Alidoust2014triplet, Fominov2015, Bernardo2015Meis, Hoshino2014Meis, Lee2017Meis, Krieger2020Meis}. This paramagnetic response has recently been experimentally confirmed  in regions where odd-frequency superconductivity is proximity-induced in various heterostructures \cite{Bernardo2015Meis, Krieger2020Meis}. In these heterostructures the paramagnetic response is not a problem, since superconductivity is originating from a conventional bulk superconductor that still has a diamagnetic Meissner effect and is thus stable in the presence of a weak magnetic field. However, finding a paramagnetic Meissner effect in the bulk of a superconductor directly raises issues of stability of the superconducting phase. In fact, the existence of a paramagnetic Meissner effect has been used as a key argument against the existence of an intrinsic bulk odd-frequency superconducting state \cite{Hoshino2014Meis, heid1995thermodynamic, Kusunose2011}.

The issue with a possible paramagnetic Meissner effect from odd-frequency pairing causes a clear conundrum for odd-frequency superconductivity in multi-orbital superconductors, as the odd-frequency pairing is here clearly a bulk effect and therefore directly raises the question of how stable these superconductors actually are. 
In this work we tackle this problem by investigating the relation between odd- and even-frequency pairing and Meissner effect in a generic two-orbital bulk superconductor, where the amount of odd-frequency pairing is directly tunable by the asymmetry in the superconducting paring between the two orbitals.
By dividing up the contributions to the Meissner effect from the even- and odd-frequency pairing and also from intra- and inter-band processes, we are able to derive simple analytical expressions that show that odd-frequency pairing generates both diamagnetic (positive) and  paramagnetic (negative) Meissner contributions. In particular, the intra- and inter-band odd-frequency contributions always have different signs, with the signs determined by the normal-state band structure. 

More specifically our results show that for a material with two electron-like bands (or hole-like), see Fig.~\ref{fig:nband}(a), we find that the odd-frequency inter-band Meissner contribution is paramagnetic, but it is always compensated by an almost equally large but diamagnetic intra-band contribution. Adding to that an overall diamagnetic contribution from the even-frequency pairing, the total Meissner effect of the superconductor is always diamagnetic, even in the limit of large odd-frequency pairing.
On the other hand, for a material with an inverted band structure, i.e.~one electron-like and one hole-like band as found in topological materials, see Fig.~\ref{fig:nband}(b), odd-frequency pairing instead generates a diamagnetic inter-band contribution to the Meissner effect that is always larger than the intra-band paramagnetic contribution. Very interestingly, in the near-metallic regime, the odd-frequency inter-band contribution becomes dominating and thus the total diamagnetic Meissner effect of the superconductor is even driven by the odd-frequency pairing. 
These results show that odd-frequency pairing in multi-orbital superconductors does  {\it not} produce a destabilizing paramagnetic Meissner effect but can even generate a diamagnetic response in topological materials.

The organization of the rest of the article is as follows.
In section \ref{sec:Theory} we first present the Hamiltonian for a generic two-orbital superconductor and then derive its superconducting pair amplitudes in subsection \ref{subsec:Model} and the general theory and calculational details for the Meissner effect in subsection \ref{subsec:Meissner}. We then present our results in section \ref{sec:results}, both analytical and numerical, divided up into two different cases: when the resulting energy bands have the same signs of their curvatures, both being electron-like in subsection \ref{res:same}, and when they are inverted, or topological, with different signs of their curvatures in subsection \ref{res:inv}). Finally, we summarize our results in section \ref{sec:con}. We also provide some more details of the calculations in Appendix \ref{app:xi12k} and \ref{app:bandbasis}.

\section{Theory}\label{sec:Theory}
\subsection{Generic two-orbital superconductor}\label{subsec:Model}

To investigate odd-frequency superconductivity and Meissner effect in multi-orbital superconductors, we assume a bulk material with a two generic low-energy orbitals in the normal state. This is the simplest model that leads to the presence of bulk odd-frequency superconducting pairing. The normal state is described in the orbital basis as $h(k)=\xi_+\tau_0+\xi_-\tau_z+\xi_{12}\tau_x$,
where $\xi_{1(2)}=\xi_{+}\pm \xi_-$ and $\xi_{12}$ represent the intra- and inter-orbital kinetic energy dispersions, respectively. Here, $\tau$ represents the Pauli matrices in the orbital basis and we assume the Hamiltonian to be spin-degenerate. Diagonalizing this normal part results in the two energy bands:
\begin{align}\label{eq:nband}
\epsilon_\pm=\xi_+\pm\sqrt{\xi_-^2+\xi_{12}^2}.
\end{align}.

\begin{figure}[!thpb]
\centering
\includegraphics[width=0.8\columnwidth]{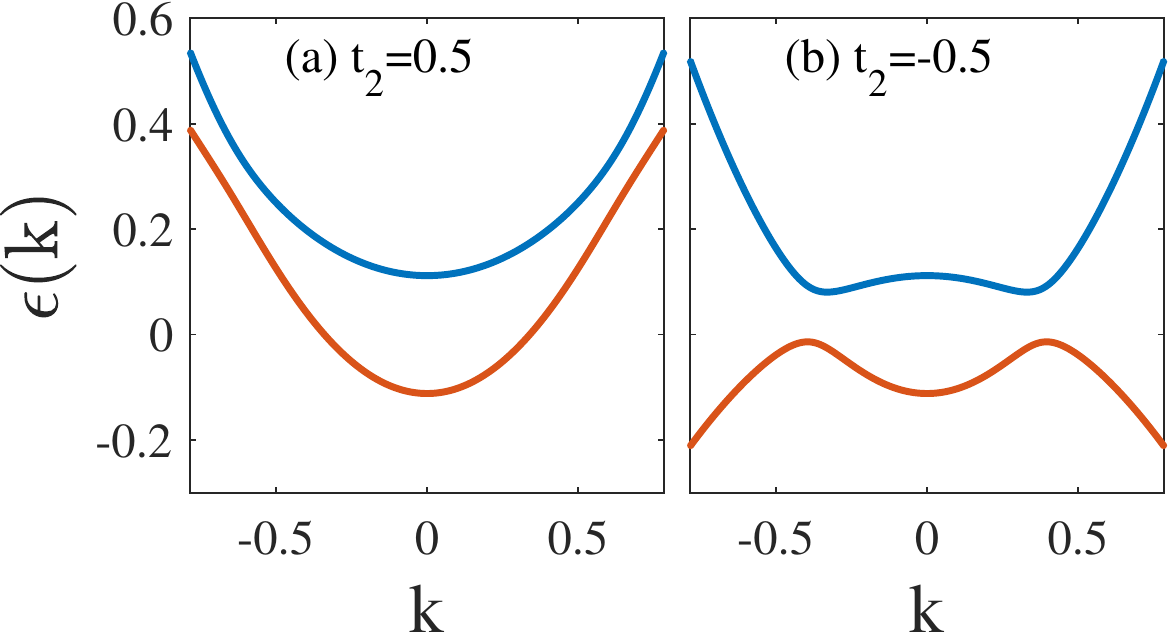}
\caption{Normal state band structure given by Eq.~\eqref{eq:nband} for a two-orbital superconductor for (a) electron-like band structure with $t_2=0.5$ and (b) inverted band structure with $t_2=-0.5$. Here $t_{12}=0.05,\mu = 0, \delta\mu=-0.2$.
\label{fig:nband}}
\end{figure}

To make our work general, yet simple enough to allow a partial analytical treatment, we assume the dispersion relations in two orbitals to be of the form $\xi_{1(2)}=-\mu+t_{1(2)}k^2-(-1)^{1(2)}\delta\mu/2$, with chemical potential $\mu$, orbital energy difference $\delta \mu$, electron wavevector $k$, and effective curvature (the inverse of the effective mass) $t_{1,2}$. For the numerical results we limit ourselves to a two-dimensional wave vector to keep the momentum integrations feasible, but we expect no significant changes for three dimensions. Moreover, we mainly assume a $k$-independent hybridization $\xi_{12}=t_{12}$ between the two orbitals, but we report complementary results for $k$-dependent hybridization $\xi_{12}(k)$  in Appendix \ref{app:xi12k}. Throughout this work, we set the energy scale through $t_1=1$ and in order to investigate the effects of the band structure, we separately treat two cases, $t_2 >0$ and $t_2 <0$.
In Fig.~\ref{fig:nband} we plot the band structure given by Eq.~\eqref{eq:nband} for the two different cases, where in (a) $t_2=0.5$ and (b) $t_2=-0.5$. Thus, in (a) we have two parabolic bands with the similar electron-like dispersions but different curvatures with their energy minima at $\pm\sqrt{t_{12}^2+(\delta\mu/2)^2}$. In contrast, in (b) we get a so-called inverted band structure, consisting originally of overlapping electron- and hole-like bands, which hybridize and form a gapped band structure, such that the Fermi level never crosses both bands. This second model has a band structure closely resembling that of topological insulators and other non-trivial topological materials.
  
We include superconductivity in the system by introducing a conventional spin-singlet intra-orbital pairing in each orbital, written as $\hat{\Delta}=(\delta_+\tau_0+\delta_-\tau_z)\otimes i \sigma_y$, with $\sigma$ being the Pauli matrices in spin space. Here the spin-singlet order parameter in each orbital is $\delta_{1(2)}=\delta_++(-)\delta_-$, which is the simplest possible implementation of superconductivity in a two-orbital model. As we will explicitly see later, odd-frequency pairing is always present in this model for a finite $\delta_-$, as long as also $\xi_{12}$ is finite. The other alternative route to generate odd-frequency pairing is to assume a finite inter-orbital pairing $\delta_{12}$ \cite{Lucia2017Kerr,Triola2020reviewAnnPhy}, but such term can always be described within a change of basis of our current model, see Appendix \ref{app:bandbasis}, and thus it cannot qualitatively change our results.
Finally, writing the full Hamiltonian in the orbital basis $\psi^\dagger=(c^\dagger_{1\uparrow} c^\dagger_{2\uparrow} c_{1\downarrow} c_{2\downarrow})$ we arrive at
\begin{align}\label{eq:Ht}
{\cal H}= \xi_+\tau_0+\xi_-\tau_z+\xi_{12}\tau_x\otimes \gamma_z + (\delta_+\tau_0+\delta_-\tau_z)\otimes \gamma_x,
\end{align}
where $\gamma$s are Pauli matrices in the particle-hole basis.
The energy eigenvalues of the full Hamiltonian, ${\cal H}$ are equal to
\begin{align}\label{eq:energy}
\varepsilon_{\pm}^2=&\xi_+^2+\xi_-^2+\xi_{12}^2+\delta_+^2+\delta_-^2
\nonumber\\
&\pm 2\sqrt{(\delta_-\delta_++\xi_-\xi_+)^2+(\delta_-^2+\xi_+^2)\xi_{12}^2}.
\end{align}

\subsubsection{Superconducting pair amplitude}
To understand superconducting pairing in our generic two-orbital superconductor, Eq.~\eqref{eq:Ht}, we extract the Green's function of the system:
\begin{align} \label{eq:GF}
 {\cal G}=(\omega-{\cal H})^{-1}\equiv \begin{pmatrix}
G&F\\ \bar{F}&\bar{G}
\end{pmatrix}.
\end{align}
Here, $G$ and $\bar{G}$ represent the electron and hole propagators, respectively, while $F$ and $\bar{F}^\dagger = F$ are the  anomalous Green's functions, or equivalently, the superconducting pair amplitude. 
The anomalous Green's function can further be decomposed into its even- and odd-frequency components $F=F^e+F^o$. For the Hamiltonian given in Eq.~\eqref{eq:Ht} we find (in the orbital basis)
\begin{align}\label{eq:Fe}
F^e=\frac{1}{D}\begin{pmatrix}
\omega^2(\delta_++\delta_-)-\alpha_+&2\xi_{12}(\delta_+\xi_+-\delta_-\xi_-)\\2\xi_{12}(\delta_+\xi_+-\delta_-\xi_-)&\omega^2(\delta_+-\delta_-)-\alpha_-
\end{pmatrix},
\end{align}
and
\begin{align}\label{eq:Fo}
F^o=\frac{1}{D}\begin{pmatrix}
0&-2\omega \xi_{12}\delta_-\\2\omega \xi_{12}\delta_-&0
\end{pmatrix}.
\end{align}
Here, $\alpha_\pm=(\delta_+\pm\delta_-)(\xi_+\mp\xi_-)^2+(\delta_+\mp\delta_-)(\xi_{12}^2+\delta_+^2-\delta_-^2)$, and denominator $D=(\omega^2-\varepsilon_+^2)(\omega^2-\varepsilon_-^2)$, thus being an even function of frequency.

The even-frequency pair amplitude in Eq.~\eqref{eq:Fe} include both intra-orbital (diagonal components) and inter-orbital (off-diagonal components) components, where the latter is directly proportional to the inter-orbital hybridization, $\xi_{12}$. The odd-frequency pair amplitude Eq.~\eqref{eq:Fo} has only inter-orbital components and is a consequence of the existence of superconducting orbital selectivity, i.e.~a finite $\delta_-$, and finite inter-orbital hybridization, $\xi_{12}$, i.e.~there must exist an asymmetry in the superconducting pairing between the two orbitals, or an orbital selectivity, together with a finite hybridization between the two orbitals for odd-frequency pairs to exist. 
In addition to generating odd-frequency pairs, the orbital selectivity term $\delta_-$ causes the superconducting part of the Hamiltonian to not commute with the normal part. This incompatibility between the normal and superconducting parts has recently also been used to define the concept of superconducting fitness \cite{Ramires2016}, which has been shown to be directly linked to the existence of odd-frequency pairs \cite{Triola2020reviewAnnPhy}. 
We can here explicitly verify that all pair amplitudes satisfy the fermionic nature of superconductivity: the inter-orbital pairs are even (odd) under the interchange of orbital index when they have an even- (odd-)frequency dependence, as required for spin-singlet pairs with no $k$-dependence ($s$-wave symmetry).  Above we explicitly worked in the orbital basis, for the form of the Hamiltonian in the band basis, where the kinetic energy is diagonal, see Appendix \ref{app:bandbasis}.

\subsection{Meissner effect}\label{subsec:Meissner}

The Meissner effect is the response of a superconductor to an externally applied magnetic field. Within linear response theory, the current response function $\textbf{j}$ to the vector potential of the magnetic field $\textbf{A}$ can be obtained via ${j}_\mu({\textbf{q}},\omega)=-K_{\mu\nu}({\textbf{q},\omega})A_\nu({\textbf{q},\omega})$. Here, $K$ is the current-current correlation function and $\nu,\mu$ are the directions of the applied vector potential and current response, respectively, while ${\textbf{q}},\omega$ are the wavevector and frequency of the response function. The Meissner response is obtained for a static and uniform magnetic field, thus taking the limits $\textbf{q}\rightarrow 0$ and $\omega \rightarrow 0$. 
To calculate $K$ we need the current operator matrix ${\cal J} = {\cal J}_\mu^p+{\cal J}_{\mu\nu}^dA_\nu$, divided into its paramagnetic and diamagnetic part and which we find in the standard way by first introducing the vector potential $\textbf{A}$ into the Hamiltonian through the substitution $\textbf{k} \rightarrow \textbf{k}-\textbf{A}$ and then taking the derivative with respect to $A_\nu$. The Meissner response can then be written as \cite{Mizoguchi2015,Johann2020}
\begin{align}\label{eq:K}
K_{\mu\nu}(\textbf{q}=0,\omega=0)=-\sum_{\omega}\sum_{{\bf k}} \mathrm{Tr_e}\left\lbrace {\cal G}{\cal J}^p_\mu{\cal G}{\cal J}^p_\nu+{\cal G}{\cal J}^d_{\mu\nu}\right\rbrace,
\end{align}
where ${\cal G}$ is the Green's function and $\mathrm{Tr_e}$ represents the trace over the electron part of the matrix. We here work in natural units, such that we set $\hbar=c=e=m=1$. Moreover, since our system is isotropic in space, we can without loss of generality focus on a specific direction, say the $x$-direction, of the Meissner response function and thus drop the $\mu,\nu$ indices. Finally, using the fact that the current operator ${\cal J}^{p(d)}$ in the particle-hole basis takes the diagonal form $\begin{pmatrix}
J^{p(d)}&0\\0&\bar{J}^{p(d)}
\end{pmatrix}$, the Meissner response in Eq.~\eqref{eq:K}, can be simplified as
\begin{align}\label{eq:Kernel}
K=-\sum_{\omega}\sum_{{\bf k}}\mathrm{Tr}\left\lbrace GJ^pGJ^p+F\bar{J}^p\bar{F}J^p+GJ^d
 \right\rbrace.
\end{align}
Since the focus of this study is to obtain the contributions of the odd- and even-frequency pairing to the Meissner effect, we only need to focus on the second term of the above equation:
\begin{align}\label{eq:KF}
K_F=-\sum_{\omega}\sum_{{\bf k}}Tr\lbrace F\bar{J}^p\bar{F}J^p \rbrace,
\end{align}
 because only this term depends on the pair amplitudes $F,\bar{F}$. Moreover, the first and the last terms usually cancel each other in conventional and also multi-band superconductors \cite{Johann2020, Asano2015}. 
 Our interest is thus focused on the Meissner Kernel of Eq.~\eqref{eq:KF}: ${\cal K}=\mathrm{Tr}\lbrace F\bar{J}^p\bar{F}J^p \rbrace$. This term can always be decomposed into ${\cal K}= {\cal K}^e + {\cal K}^o$, where ${\cal K}^e=\mathrm{Tr}\lbrace F^e\bar{J}^p\bar{F}^eJ^p \rbrace$ is entirely due to the even-frequency pairing, while ${\cal K}^o=\mathrm{Tr}\lbrace F^o\bar{J}^p\bar{F}^oJ^p \rbrace$ comes exclusively from the odd-frequency pairing. 
Technically there also exists terms of the form $F^o\bar{J}^p\bar{F}^eJ^p$ in ${\cal K}$, but, since they are always odd in frequency, they automatically cancel during the final frequency summation in Eq.~\eqref{eq:Kernel}. 

\subsubsection{Kernel decomposition into intra- and inter-band processes}
To better understand the physics of the Meissner response, we decompose its Kernel ${\cal K}$ into parts coming from inter- and intra-band processes, respectively. Using the fact that the Meissner Kernel can generically be written as ${\cal K}=(a\omega^4+2b\omega^2+c)/D^2$, we can form the following decomposition:
\begin{align}\label{eq:Kernel-decomp}
{\cal K}=\frac{{\cal K}_+}{(\omega^2-\varepsilon_+^2)^2}+\frac{{\cal K}_-}{(\omega^2-\varepsilon_-^2)^2}
+\frac{{\cal K}_{12}}{(\omega^2-\varepsilon_+^2)(\omega^2-\varepsilon_-^2)}.
\end{align}
Here ${\cal K}_{\pm}$ and ${\cal K}_{12}$ are frequency-independent coefficients for processes involving only one individual band (i.e.~intra-band processes) and both bands (i.e.~inter-band processes), respectively. These coefficients are given by
\begin{align}\label{eq:Kerneld}
{\cal K}_{\pm}&=\frac{a\varepsilon_{\pm}^4+2b\varepsilon_{\pm}^2+c}{(\varepsilon_+^2-\varepsilon_-^2)^2}\nonumber\\
{\cal K}_{12}&=-2\frac{a\varepsilon_{+}^2\varepsilon_-^2+b(\varepsilon_{+}^2+\varepsilon_-^2)+c}{(\varepsilon_+^2-\varepsilon_-^2)^2}.
\end{align} 
We here note that the intra-band term is the only term appearing in a single-orbital superconductor, or for a two-orbital superconductor in the trivial $\xi_{12}=0$ limit, while the inter-band term only appears when $\xi_{12}$ is finite. We also emphasize here that these are processes between different {\it bands}, i.e.~the eigenstates of the Hamiltonian Eq.~\eqref{eq:Ht}, and not processes between the different orbitals.

\subsubsection{Frequency summation}
In order to analytically perform the summation over frequency in the Meissner Kernel in Eq.~\eqref{eq:KF}, we use analytical continuation from real frequency to Matsubara frequency, $\omega \rightarrow i\omega_n$.
Then, by applying Matsubara frequency summation, the Meissner response reads
\begin{align}\label{eq:KF2}
K_F=\sum_{\bf k} &\frac{{\cal K}_+}{2 \varepsilon_+^2} (\mathcal{C}(\varepsilon_+)+n^\prime(\varepsilon_+))+\frac{{\cal K}_-}{2 \varepsilon_-^2} (\mathcal{C}(\varepsilon_-)+n^\prime(\varepsilon_-))\nonumber\\
+&{\cal K}_{12}\frac{\mathcal{C}(\varepsilon_-) - \mathcal{C}(\varepsilon_+)}{\varepsilon_+^2 - \varepsilon_-^2},
\end{align}
where $n(\xi)$ is the Fermi-Dirac distribution function, $n'(\xi)$ its derivative with respect to energy, and ${\cal C}(\xi)=(n(-\xi)-n(\xi))/2\xi$. 
Here we note that both the ${\cal C}(\varepsilon)$ and $\frac{\mathcal{C}(\varepsilon_-) - \mathcal{C}(\varepsilon_+)}{\varepsilon_+^2 - \varepsilon_-^2}$ functions are always positive, while $n^\prime(\varepsilon)$ is always negative. Moreover, assuming we always have a gapped superconductor, as is usually the case for the bulk of $s$-wave states, the $n^\prime(\varepsilon)$ term is negligible, and hence it is the signs of ${\cal K}_\pm$ and ${\cal K}_{12}$ that become the decisive factors for determining the sign of the Meissner effect.

We can further simplify the expression for the Meissner effect by noting that, if we have a $k$-independent orbital hybridization $\xi_{12}$ as we assume here in the main text, the current operator is diagonal in the orbital basis: $J^p=\begin{pmatrix}
j_1&0\\0&j_2
\end{pmatrix}$ and similarly for $\bar{J}^p=-J^p$ (very results for a $k$-dependent hybridization are reported in Appendix \ref{app:xi12k}).
Moreover, considering that the anomalous Green's function takes the form $\begin{pmatrix}
f_1&f_{12}\\f_{21}&f_2
\end{pmatrix}$, the Meissner Kernel can be written 
\begin{align}\label{eq:simpleK}
{\cal K}=j_1^2f_1^2+j_2^2f_2^2+2j_1j_2f_{12}f_{21} = (a\omega^4+2b\omega^2+c)/D^2,
\end{align}
where the last equality is just iterating the statement from before used to parametrize the Meissner Kernel.
 Here, the first two terms in the first expression are always positive, while the last term changes sign based on the sign of $j_1j_2$ and $f_{12}f_{21}$. Since $j_{1(2)}=-\partial \xi_{1(2)}(k-A)/\partial A =2t_{1(2)}k$, the sign of the product $j_1j_2$ is equivalent to the sign of $t_2$, since we set $t_1=1$. It is thus natural when studying the Meissner effect to distinguish between the two cases: $t_2>0$ giving $j_1j_2>0$, which has two bands with the same curvature, and $t_2<0$ with $j_1j_2<0$, which has an inverted band structure. This is particularly important for the case of the odd-frequency pairing contribution to the Meissner effect as this pairing contains only the interorbital terms $f_{12},f_{21}$.

\section{Results} \label{sec:results}
Having established the underlying theory we can now focus on calculating the Meissner effect in generic two-orbital superconductors described by Eq.~\eqref{eq:Ht}. Thanks to the derivations in the preceding section we can proceed analytically to a large degree. To calculate the Meissner Kernel in terms of its intra- and inter-band contributions in Eq.~\eqref{eq:Kernel-decomp}, we need the coefficients $a,b,c$ in Eq.~\eqref{eq:Kerneld}. For a $k$-independent orbital hybridization $\xi_{12}$ these are directly accessible using Eq.~\eqref{eq:simpleK}. 
For the contribution to the Meissner Kernel from the even-frequency pairing, we use Eq.~\eqref{eq:Fe} for the anomalous Green's function components and arrive at the coefficients:
\begin{align}\label{eq:abcKeven}
a^e&=j_1^2\delta_1^2+j_2^2\delta_2^2\nonumber\\
b^e&=-(j_1^2\alpha_+\delta_1+j_2^2\alpha_-\delta_2)\nonumber\\
c^e&=j_1^2\alpha_+^2\delta_1+j_2^2\alpha_-^2+2j_1j_2f_{12}^2.
\end{align}
These, plugged in Eq.~\eqref{eq:Kerneld}, directly give the even-frequency Meissner Kernel contributions ${\cal K}_{\pm}^e$ and ${\cal K}_{12}^e$. 
For the contributions to the Meissner Kernel from the odd-frequency pairing we instead use Eq.~\eqref{eq:Fo} and find that $a^o=c^o=0$, which means we can straightforwardly simplify the expressions to arrive at
\begin{align}\label{eq:Ko-simp}
{\cal K}^o_\pm&=j_1j_2\frac{8\delta_-^2\xi_{12}^2\varepsilon_\pm^2}{(\varepsilon_+^2-\varepsilon_-^2)^2}\nonumber\\
{\cal K}^o_{12}&=-j_1j_2\frac{8\delta_-^2\xi_{12}^2(\varepsilon_-^2+\varepsilon_+^2)}{(\varepsilon_+^2-\varepsilon_-^2)^2}.
\end{align}
These equations show that for odd-frequency pairing, the sign of the intra- and inter-band Meissner Kernels, and thus their contributions to the Meissner effect, are opposite to each other and their sign only depends on the sign of $j_1j_2$. Thus, for a band structure with two bands with the same curvature, odd-frequency pairing always has a positive or diamagnetic Meissner effect from intra-band processes, while inter-band processes always give a negative or paramagnetic Meissner effect.
On the other hand, for an inverted band structure, where then $j_1j_2<0$, we get the opposite behavior: intra-band processes always generate a paramagnetic Meissner effect, while inter-band processes always give a diamagnetic effect.

To further analyze the results, especially establishing the relative sizes of the even- and odd-frequency and intra- and inter-band contributions to the Meissner effect, we need to perform the summation over reciprocal space in Eq.~\eqref{eq:KF2}.  
Having established the importance of the sign of the product $j_1j_2$ in Eq.~\eqref{eq:simpleK}, we divide our numerical results into two distinct cases: two electron-like bands where $j_1j_2>0$ (or equivalently two hole-like bands) and an inverted band structure with $j_1j_2<0$. In particular, below we choose $t_2 = \pm 0.5$ and also set $\delta_+=0.03$ and $\xi_{12}$ to an $k$-independent constant $t_{12}$. We then vary all other parameters; $\mu$, $\delta \mu$, and $\delta_-$, to arrive at a comprehensive picture of the Meissner effect in generic two-orbital superconductors. In terms of determining the impact of odd-frequency pairing on the Meissner effect, it is particularly important to vary $\delta_-$ as the odd-frequency pair amplitude is directly proportional to this pairing asymmetry, or orbital selectivity.

\subsection{Two electron-like bands, $j_1j_2>0$}\label{res:same}
We begin with a material consisting of two low-energy electron-like bands, here generically modeled with $t_1 =1$ and $t_2=0.5$, such that $j_1j_2>0$.
In this case the odd-frequency pairing generates an intra-band Meissner effect that is always diamagnetic, while the inter-band part is always paramagnetic, as given by Eq.~\eqref{eq:Ko-simp}. However, as we in the plots see below, even-frequency pairing generate terms which can change sign depending on parameter values. 

\begin{figure}[!thpb]
\centering
\includegraphics[width=0.95\linewidth]{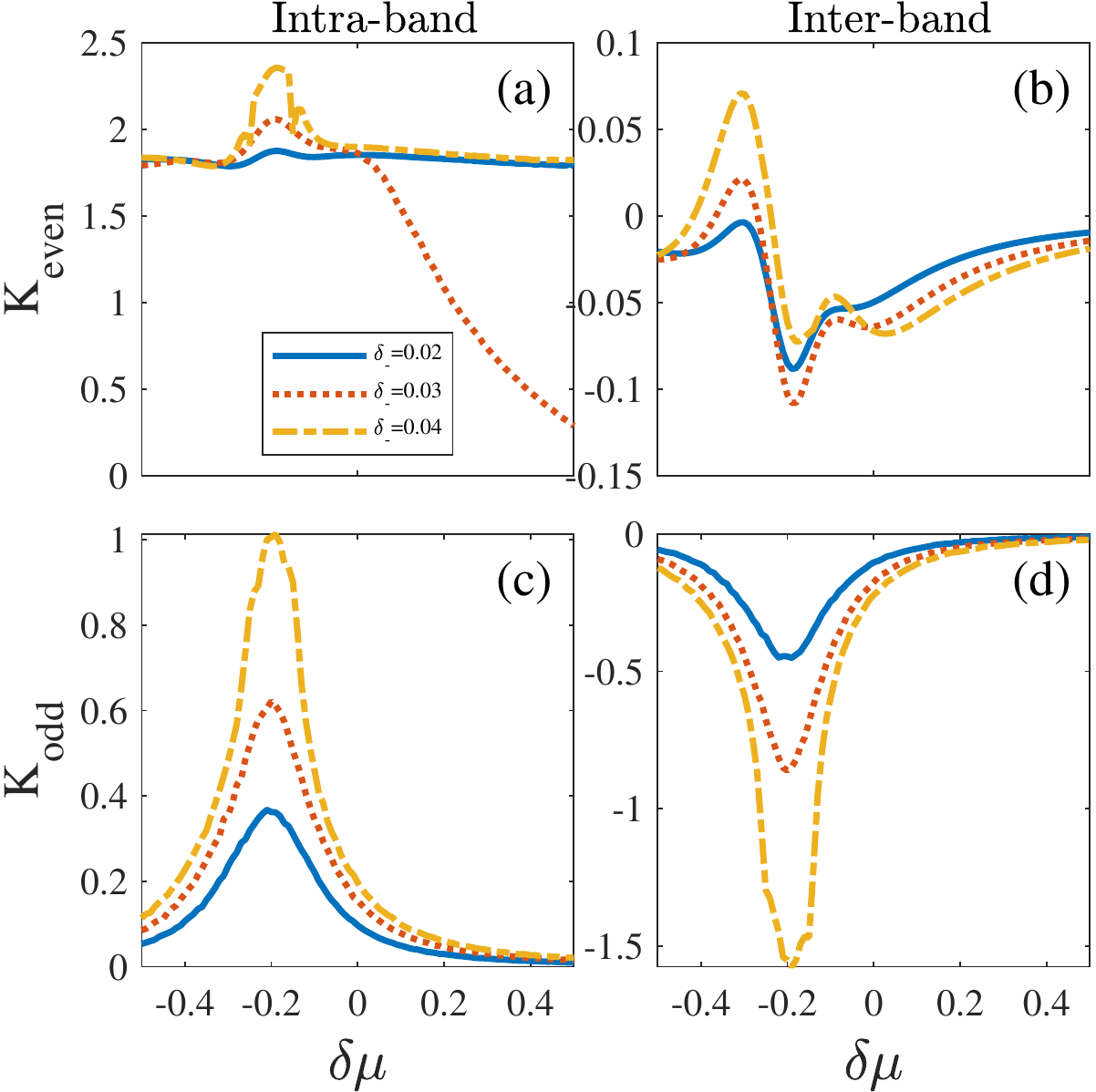}
\caption{Meissner effect divided into even- and odd-frequency (rows) and intra- and inter-band (column) contributions as a function of energy difference between orbitals $\delta\mu$ and for several values of orbital pairing asymmetry $\delta_-$. Here $t_2=0.5$, $t_{12}=0.05$, $\mu=0.3$, and $\delta_+=0.03$.
\label{fig:K-dmu-t2-plus}}
\end{figure}

We start by investigating in Fig.~\ref{fig:K-dmu-t2-plus} the even- and odd-frequency contributions to the Meissner effect divided up into intra- and inter-band processes and as a function of the energy difference between orbitals $\delta\mu$, for different values of orbital selective superconducting order parameter $\delta_-$. As we fix $\delta_+=0.03$, we here investigate three different regimes. For $\delta_+>\delta_-$ we are in a so-called $s_{++}$ phase where the phase of the superconducting order parameter on each orbital is the same: $\delta_{1,2}>0$, while for $\delta_+<\delta_-$ we are in a $s_{+-}$ phase where $\delta_1>0$ but $\delta_2<0$. At the boundary, $\delta_-=\delta_+=0.03$, we find $\delta_2 =0$, i.e.~only orbital 1 is intrinsically superconducting. Moreover, we set $\mu=0.3$ which makes the Fermi level cross both bands, see Fig.~\ref{fig:nband}(a), and thus both bands give large contributions to the Meissner effect.

In Figs.~\ref{fig:K-dmu-t2-plus}(a,b) we plot the contributions from the even-frequency pairing to the Meissner response. The contribution from intra-band processes (a) is completely dominating and remain positive for the full range of parameters values, i.e.~a diamagnetic Meissner response. The behavior is more or less the constant for both $s_{++}$ and $s_{+-}$ orbital pairing. The only exception is the special case $\delta_+=\delta_-$, which developes a notable decrease when $\delta\mu$ is increasing. We can understand this behavior by noting that, in this case there is no intrinsic superconductivity in orbital 2 ($\delta_2=0$) and increasing $\delta\mu$ shifts the band bottom of orbital 1 to higher energies, thus making its Fermi surface smaller. As a consequence, the Meissner intra-band contributions  goes down with increasing  $\delta\mu$. While the even-frequency inter-band contribution in (b) is much smaller, we note that it interestingly has no fixed sign. 
Turning to the odd-frequency contributions, we see in Figs.~\ref{fig:K-dmu-t2-plus}(c,d) that the intra- (inter-)band part is dia- (para)magnetic, as also established by Eq.~\eqref{eq:Ko-simp}. Furthermore, we find that the inter-band contribution is generally larger than the intra-band contribution, in contrast to the dominating intra-band process for the even-frequency pairing. This leaves the total Meissner effect from odd-frequency pairing paramagnetic, in agreement with the traditional expectation for the odd-frequency response \cite{Abrahams1995, Yokoyama2011, Alidoust2014triplet, Fominov2015, Bernardo2015Meis, Hoshino2014Meis, Lee2017Meis, Krieger2020Meis}, but we note that the overall effect is partly canceled due to a relatively large diamagnetic intra-band contribution. Moreover, we find that the Meissner response from the odd-frequency pairing is significantly increasing with increasing orbital asymmetry $\delta_-$, in agreement with the effectively linear dependence on $\delta_-$ for the odd-frequency pair amplitude.

\begin{figure}[!thpb]
\centering
\includegraphics[width=0.95\linewidth]{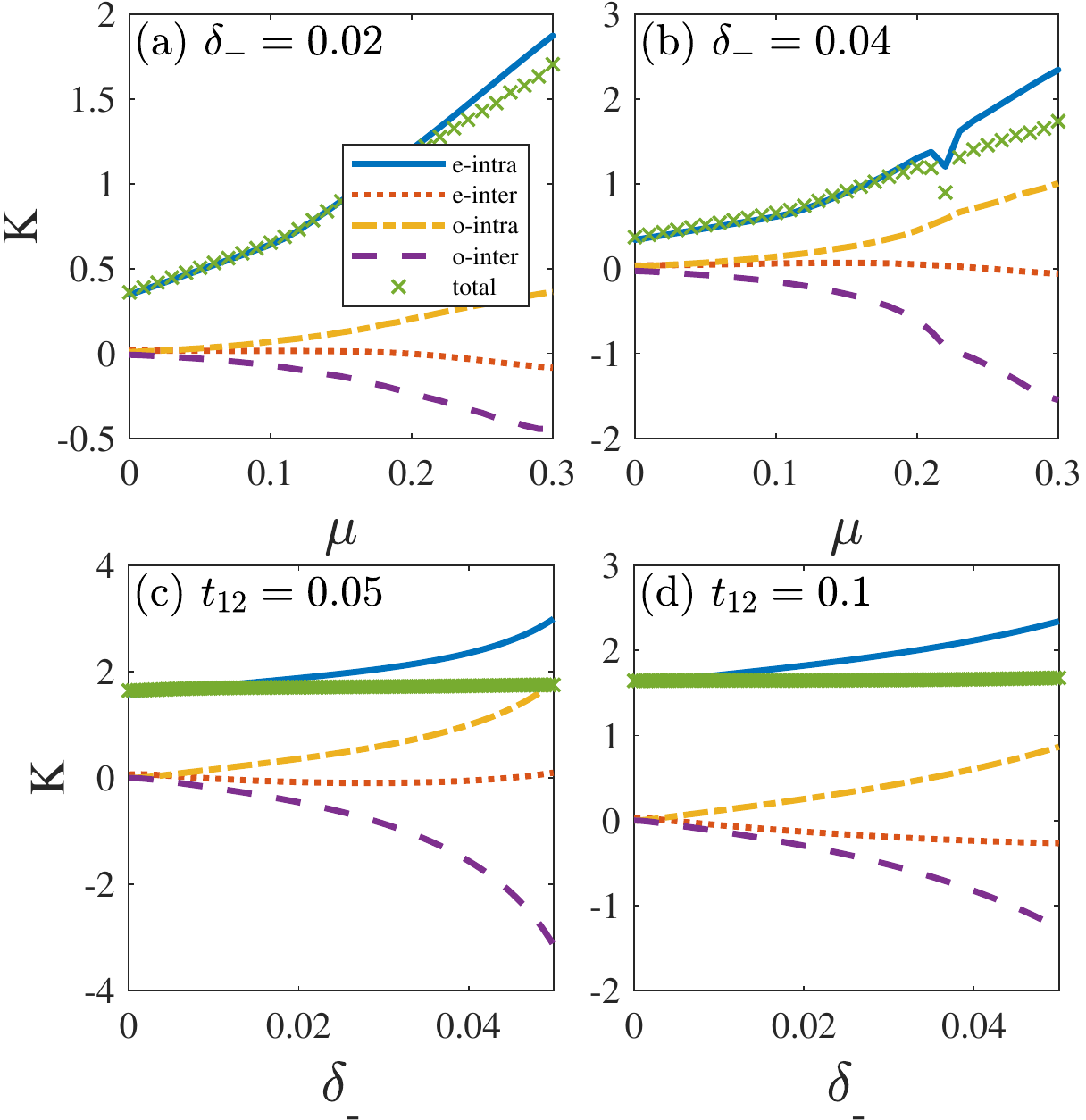}
\caption{Meissner effect as a function of overall chemical potential $\mu$ (a,b) and superconducting orbital asymmetry $\delta_-$ (c,d) divided into even-frequency intra-band (blue solid), even-frequency inter-band (red dotted), odd-frequency intra-band (yellow dash-dotted), and odd-frequency inter-band (purple dashed) contributions, together with the total effect (green $\times$ marks). Here $t_2=0.5$, $\delta\mu=-0.2$, $\delta_+=0.03$, and $t_{12}=0.05$ (a,b) and $\mu=0.3$ (c,d), with remaining parameters given in the figure panels.
\label{fig:Ktplusmudeltam}}
\end{figure}
In Fig.~\ref{fig:Ktplusmudeltam} we continue to analyze the Meissner response, focusing especially on how the different components compare to each other, with even-frequency intra-band (blue), even-frequency inter-band (red), odd-frequency intra-band (yellow), and odd-frequency inter-band (purple) together with the total response (green). Based on the distinct behaviors observed in Fig.~\ref{fig:K-dmu-t2-plus}, we here focus on the overall dependence on the chemical potential (a,b), and the superconducting asymmetry $\delta_-$, as they determine the Fermi surface size and magnitude of the odd-frequency pairing, respectively.
For the Meissner effect as a function of chemical potential we choose to report values for both the $s_{++}$ (a) and $s_{+-}$ (b) regimes. We here see how all terms increase  as a function of $\mu$, which is due to the Fermi surface increasing with $\mu$ and thus creating a stronger superconducting state, which is then reflected in an increase in the total Meissner effect.
The only minor exception to this increasing trend is the even-frequency intra-band contribution around $\mu \sim 0.22$ for  $s_{+-}$ pairing. Analyzing the superconducting order parameters in the band basis, we find that one of the intra-band order parameters $\delta_+-\delta_-\cos(\theta)$ becomes zero at this particular point, thus causing the drop in the intra-band Meissner effect (for details and definition of $\theta$, see Appendix \ref{app:bandbasis}). This drop does not exist in (a) because there $\delta_+>\delta_-$ and hence $\delta_+-\delta_-\cos(\theta)$ never becomes zero.

We also find in Fig.~\ref{fig:Ktplusmudeltam}(a,b) that both the intra- and inter-band odd-frequency contributions become larger when $\delta_-$ increases. This increase is further verified in Fig.~\ref{fig:Ktplusmudeltam}(c,d), where we plot the different contributions to the Meissner effect as a function of the superconducting asymmetry $\delta_-$. Here, we choose two different values of the inter-orbital hybridization $t_{12}=0.05$ (c) and $0.1$ (d), with all other parameters similar to Fig.~\ref{fig:Ktplusmudeltam}(a,b). All Meissner contributions increase as a function of $\delta_-$, but notably the odd-frequency terms increases faster. This is to be expected as the odd-frequency pair amplitude is linearly proportional to $\delta_-$, see Eq.~\eqref{eq:Fo}. Note, however, that we do not find that the odd-frequency Meissner contributions increase in the same manner for the orbital hybridization $t_{12}$, despite this term also being necessary to generate odd-frequency pair amplitude $F^o$, see Eq.~\eqref{eq:Fo}. This is due to the term $t_{12}$ also changing the whole band structure and thus the pair amplitude (through the denominator $D$) and the Meissner response are influenced rather substantially. We find numerically a maximum odd-frequency Meissner response around $t_{12}\sim 0.05$, see Appendix \ref{app:xi12k}. With $\delta_-$ being a much smaller energy scale, the same band structure effects are not seen for realistic $\delta_-$.
Interestingly, we find that for the odd-frequency part, the inter-band contributions are larger the intra-band ones for all parameters in Fig.~\ref{fig:Ktplusmudeltam}. However, the difference is always small, so, although the inter-band contribution is paramagnetic, the diamagnetic intra-band contribution makes the total paramagnetic Meissner effect from the odd-frequency pairing quite small. 
 As a consequence, odd-frequency pairing has only a minor effect on the Meissner effect in two-orbital superconductors with two electron-like (or hole-like) bands. In fact, we find that the total Meissner effect is essentially constant when $\delta_-$ increases, despite this causing a strong increase in both odd-frequency pair amplitude and its influence on the Meissner effect. Hence we conclude that multi-orbital superconductors are highly stable in a magnetic field, even when substantial bulk odd-frequency pairing is present. We emphasize that this is not due to small contributions to the Meissner effect from the odd-frequency pairing, but the fact that the intra- and inter-band odd-frequency processes nearly cancel each other. 

\subsection{Two inverted bands, $j_1j_2<0$}\label{res:inv}
We next turn to the case of an inverted band structure, such that  $t_1t_2<0$, and equivalently $j_1j_2<0$, as schematically illustrated in Fig.~\ref{fig:nband}(b) and typical for topological insulators \cite{Hasan2010}. This band structure never has both bands crossing the Fermi level, and there also needs to be a finite doping to reach beyond the insulating state.
In this case, Eq.~\eqref{eq:Ko-simp} shows that the odd-frequency pairing always gives a paramagnetic Meissner effect for intra-band processes and a diamagnetic Meissner effect for the interband processes, i.e.~opposite to the behavior in the previous section with a non-inverted band structure. This analytical result is in agreement with what has been found earlier for nematic inter-orbital spin-triplet superconductivity in doped topological insulators \cite{Johann2020}, indicating that these signs of the Meissner effect contributions are likely stable for a range of different symmetries for $\hat{\Delta}$ even though we here concentrate on intra-orbital spin-singlet $s$-wave symmetry.

\begin{figure}[!thpb]
\centering
\includegraphics[width=0.95\linewidth]{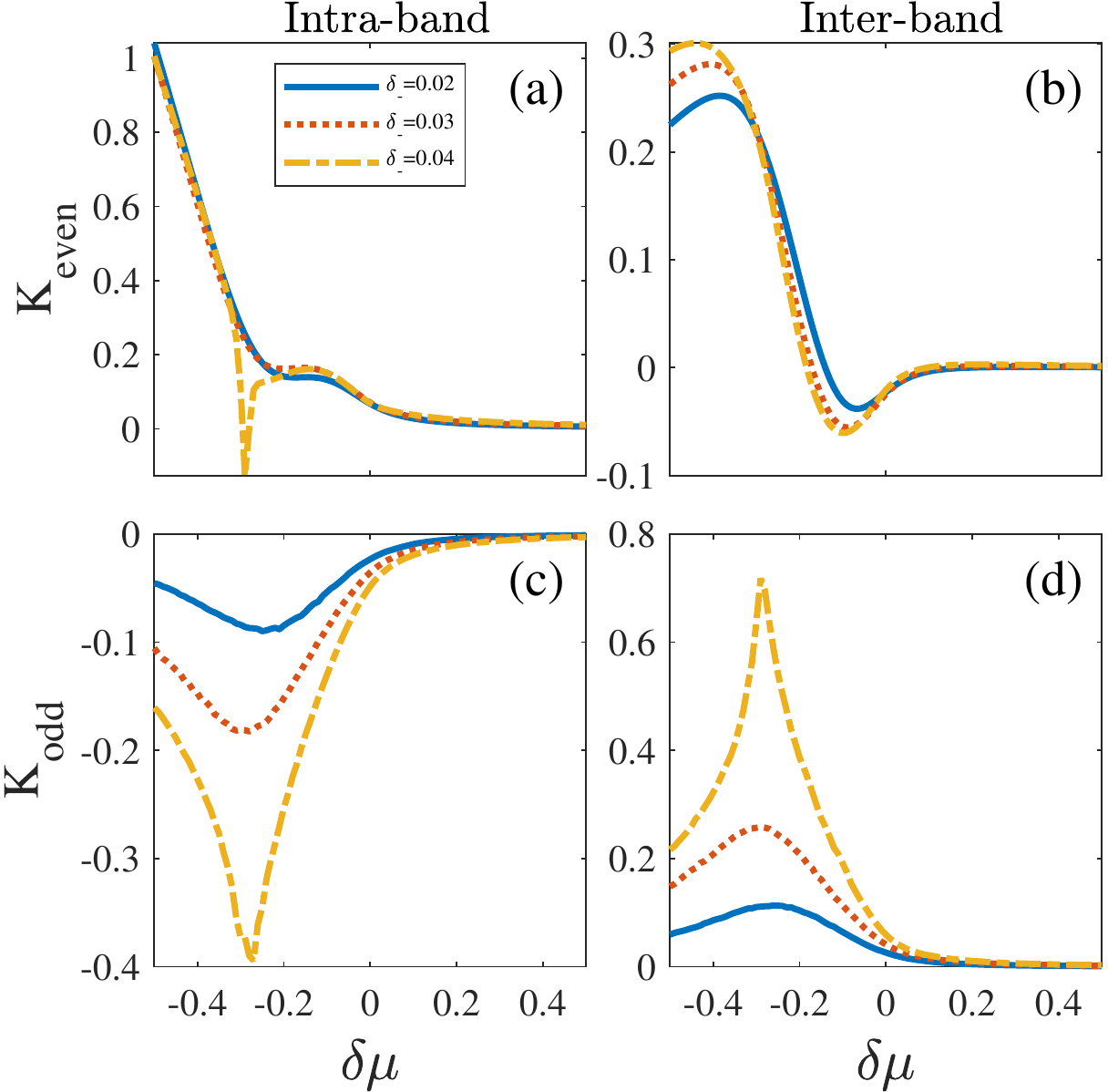}
\caption{Meissner effect divided into even- and odd-frequency (rows) and intra- and inter-band (columns) contributions as a function of energy difference between orbitals $\delta\mu$ and for several values of orbital pairing asymmetry $\delta_-$. Here $t_2=-0.5$, $t_{12}=0.05$, $\mu=0$, and $\delta_+=0.03$.
\label{fig:K-dmu-t2-minus}}
\end{figure}

Proceeding with numerical results, we next report similar plots for the inverted band structure as in for two electron-like bands in Figs.~\ref{fig:K-dmu-t2-plus}-\ref{fig:Ktplusmudeltam}.
Thus, in Fig.~\ref{fig:K-dmu-t2-minus} we show the even- and odd-frequency (rows) and intra- and inter-band (columns) contributions to the Meissner effect as a function of $\delta\mu$. We here initially fix the chemical potential $\mu =0$ such that both bands contribute similarly at $\delta \mu =0$, but we check the $\mu$ dependence in the next figure. Again we choose three increasing values of  $\delta_-$ creating $\delta_1\delta_2>0$ ($s_{++}$), $\delta_2=0$, and $\delta_1\delta_2>0$ ($s_{+-}$) orbital pairing, respectively. 
Here we find that all Meissner contributions increases rapidly when $\delta \mu$ becomes more negative until around $\delta \mu \sim -0.3$. This is the value when the Fermi level crosses into the valence band, and thus this increase is due to more low-energy states being available. For even more negative values of $\delta \mu$ we see different trends. The even-frequency intra-band contribution (a) continues to increase, as expected for a metallic state with an increasing Fermi surface. The only exception is the case of $\delta_-=0.04$, where, in a very limited regime, we surprisingly find a negative, paramagnetic contribution. This is the regime where superconductivity in one of the bands, $\delta_+-\delta_-\cos(\theta)$, approaches zero, see Appendix \ref{app:bandbasis}. The even-frequency inter-band contribution (b) gives both diamagnetic and paramagnetic responses, dependent on the choice of parameters, and is now notably larger than in the previous case with $j_1j_2>0$ in Fig.~\ref{fig:K-dmu-t2-plus}. However, for the even-frequency contribution the intra-band processes are still generally dominating.
When it comes to the odd-frequency contributions to the Meissner effect, we see how they form a peak structure quite similarly to that of Fig. ~\ref{fig:K-dmu-t2-plus}, indicating a general behavior of the odd-frequency contributions. Most importantly, here the diamagnetic inter-band contribution is much larger than the paramagnetic intra-band contribution, which leaves the odd-frequency pairing giving a diamagnetic contribution to the total Meissner response. Thus, odd-frequency pairing actually helps stabilizing the superconducting state through the generation of an increased diamagnetic Meissner effect, in complete contrast to many other systems where an odd-frequency paramagnetic Meissner effect has often been discussed \cite{Abrahams1995, Yokoyama2011, Alidoust2014triplet, Fominov2015, Bernardo2015Meis, Hoshino2014Meis, Lee2017Meis, Krieger2020Meis}.

\begin{figure}[!thpb]
\centering
\includegraphics[width=0.95\linewidth]{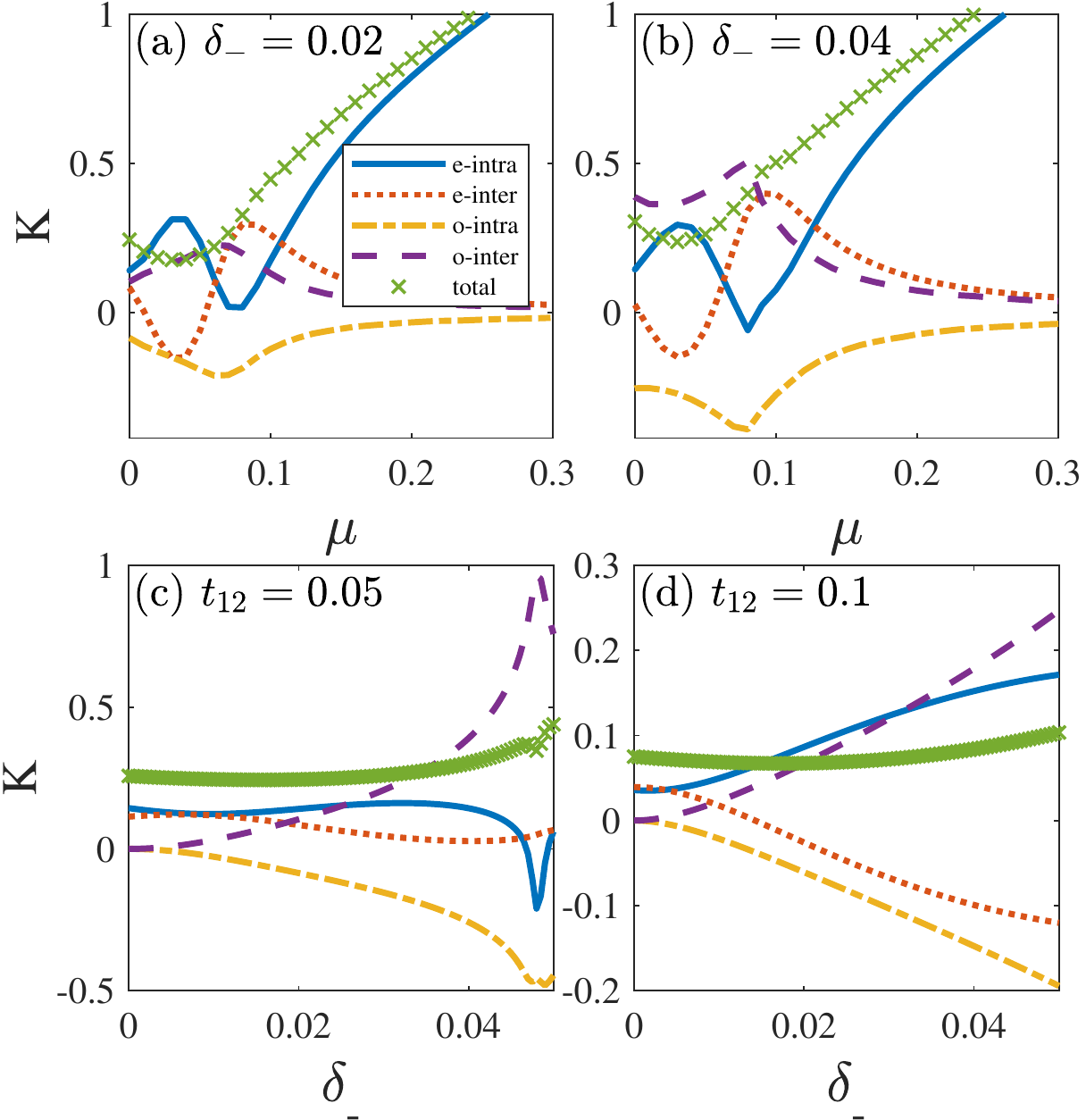}
\caption{Meissner effect as a function of overall chemical potential $\mu$ (a,b) and superconducting orbital asymmetry $\delta_-$ (c,d) 
divided into even-frequency intra-band (blue solid), even-frequency inter-band (red dotted), odd-frequency intra-band (yellow dash-dotted), and odd-frequency inter-band (purple dashed) contributions, together with the total effect (green $\times$ marks). Here $t_2=-0.5$, $\delta\mu=-0.2$, $\delta_+=0.03$, and $t_{12}=0.05$ (a,b) and $\mu=0$ (c,d), with remaining parameters given in the figure panels.
\label{fig:K-mudelm-t2-minus}}
\end{figure}

Finally, in Fig.~\ref{fig:K-mudelm-t2-minus} we plot in the same subpanels all contributions to the Meissner effect as a function of chemical potential $\mu$ (a,b) and superconducting orbital asymmetry $\delta_-$ (c,d), i.e.~similarly to Fig.~\ref{fig:Ktplusmudeltam} but now for the inverted band structure. 
For small values of the chemical potential the system is an insulator in the normal state and we find that all contributions to the Meissner effect are often of similar size. 
Interestingly, as we increase  $\delta_-$ by comparing (a,b), we get a notably higher contribution from the odd-frequency pairing and the odd-frequency inter-band part even becomes the clearly dominant contribution in this low $\mu$ regime. 
However, by increasing the chemical potential to larger values, the conventional even-frequency intra-band part increases, and eventually becomes dominant, with the total Meissner response now instead approaching its value. The transition between these two regimes is marked by the system going into a metallic regime around $\mu\sim 0.08$. It is thus not surprising that inter-band processes are dominating in the insulating and near-metallic regime, $\mu < 0.08$.
Thus we find that the odd-frequency contributions to the Meissner effect is completely dominating in the insulating and near-metallic regimes. Notably, the total Meissner effect is always diamagnetic, even with dominating odd-frequency contributions as they are also diamagnetic.

To further investigate the behavior of the Meissner effect when it is dominantly coming from the odd-frequency pairing, we plot in Fig.~\ref{fig:K-mudelm-t2-minus}(c,d) the Meissner contributions as a function of $\delta_-$ for $\mu =0$.
These subpanels clearly show how, by increasing the $\delta_-$ term both the intra- and inter-band contributions to the odd-frequency Meissner effect become large, especially for moderate $t_{12}\sim 0.05$ (c). While the intra-band contribution is paramagnetic, it is much smaller than the diamagnetic inter-band contribution, which makes the total Meissner response follow closely that of the odd-frequency inter-band contribution.
Based on these results, we conclude that for inverted, or topological, band structures, odd-frequency pairing can easily dominate the Meissner response, especially at low doping levels in the near-metallic regime. It is inter-band odd-frequency processes that dominate in this regime and they always contribute diamagnetically to the Meissner effect. As a consequence, it is only the large odd-frequency pairing that stabilizes the superconducting state in a magnetic field.

\section{Conclusions}\label{sec:con}
Odd-frequency pairing has often been discussed to give a paramagnetic Meissner effect, with only conventional even-frequency pairing assume to give the diamagnetic Meissner effect necessary to stabilize a superconductor in an external magnetic field \cite{Abrahams1995, Yokoyama2011, Alidoust2014triplet, Fominov2015, Bernardo2015Meis, Hoshino2014Meis, Lee2017Meis, Krieger2020Meis}. As a result, the existence of large odd-frequency pair amplitudes has been thought to destabilize superconductivity, possibly even to the degree where it cannot exist as a bulk effect \cite{Hoshino2014Meis}. Here we disprove this simplistic picture for multi-orbital superconductors where odd-frequency bulk superconducting pairing is ubiquitous. We show that odd-frequency pairing actually generates both dia- and paramagnetic contributions, and that they usually either nearly cancel each other or even generate an overall diamagnetic Meissner effect. As a result, even large odd-frequency pairing in the bulk does not destabilize superconductivity in multi-orbital superconductors.

In more detail, we are able to derive, using only a few non-restrictive assumptions, simple analytical results for the Meissner response in a generic two-orbital superconductor, divided up into its contributions from even- and odd-frequency pairing, as well as coming from intra- and inter-band processes. In a two-orbital superconductor, odd-frequency pairing is always present as soon as there exists a finite orbital hybridization and an asymmetry in the superconducting pairing between the two orbitals, with the odd-frequency pair amplitude growing linearly with the latter. 
We show analytically that for odd-frequency pairing, intra- and inter-band processes contribute with different signs to the Meissner effect, with the signs directly tied to the character of the normal state band structure. 

For two electron-like (or hole-like) bands (i.e.~same sign of the curvature in both bands) we always find that the odd-frequency pairing gives diamagnetic intra-band but paramagnetic inter-band contributions to the Meissner effect. Numerically we further show that, while this inter-band part is usually slightly larger,  the two contributions nearly cancel in all relevant parameter regimes. Together with a relatively large even-frequency diamagnetic Meissner response, this results in an overall stabilizing diamagnetic Meissner effect, even in the limit of large odd-frequency pairing in the superconductor.
In contrast, for an inverted band structure (i.e.~with different signs of the curvatures in the two bands, as found in topological materials), we instead always find that the odd-frequency pairing gives diamagnetic inter-band but paramagnetic intra-band contributions. Very interestingly, the odd-frequency inter-band contribution becomes even the dominating contribution to the total Meissner effect in the near-metallic regime where the Fermi level is close to the valence or conduction band bottom. Thus, the Meissner effect can even be completely driven by the odd-frequency pairing, but still be diamagnetic.
Moreover, for both types of band structures we find that the even-frequency contributions to the Meissner effect is often diamagnetic but can actually become paramagnetic for some parameter ranges. Still, we always find a stable diamagnetic Meissner effect, even in superconductors with large odd-frequency bulk pairing. 

To summarize, our results show that bulk odd-frequency bulk pairing does not cause a destabilizing paramagnetic Meissner effect in multi-orbital superconductors. Instead, odd-frequency pairing either does not contribute significantly to the Meissner effect or, very interestingly, gives a diamagnetic contribution in topological band structures. 
Based on this diamagnetic odd-frequency Meissner response in topological materials, we speculate that odd-frequency pairing might be quite common in a wide range of topological materials.

\acknowledgments{The authors thank J.~Schmidt for fruitful discussions and acknowledge financial support from the European Research Council (ERC) under the European Unions Horizon 2020 research and innovation programme (ERC-2017-StG-757553) and the Knut and Alice Wallenberg Foundation through the Wallenberg Academy Fellows program.}

\appendix

\section{Momentum dependent inter-orbital hybridization}\label{app:xi12k}
An assumption made in the main text was to assume a constant hybridization between the two orbitals: $\xi_{12}= t_{12}$. This offers a significant simplification as it gives a current operator of the form $\begin{pmatrix}
j_1&0\\0&j_2
\end{pmatrix}$, which is needed in order to proceed with the analytical calculations. Without this assumption we could not derive the relatively simple expressions in Eqs.~\eqref{eq:abcKeven}-\eqref{eq:Ko-simp} for the Meissner effect from even- and odd-frequency pairing.
 In this appendix, we first present the behavior of Meissner effect with respect to $t_{12}$ in order to provide a comprehensive view of the dependence on the orbital hybridization, and then we present results for a momentum dependent hybridization $\xi_{12}(k)$, showing that such a change does not qualitatively change our conclusions. 

\begin{figure}[!thpb]
\centering
\includegraphics[width=0.95\linewidth]{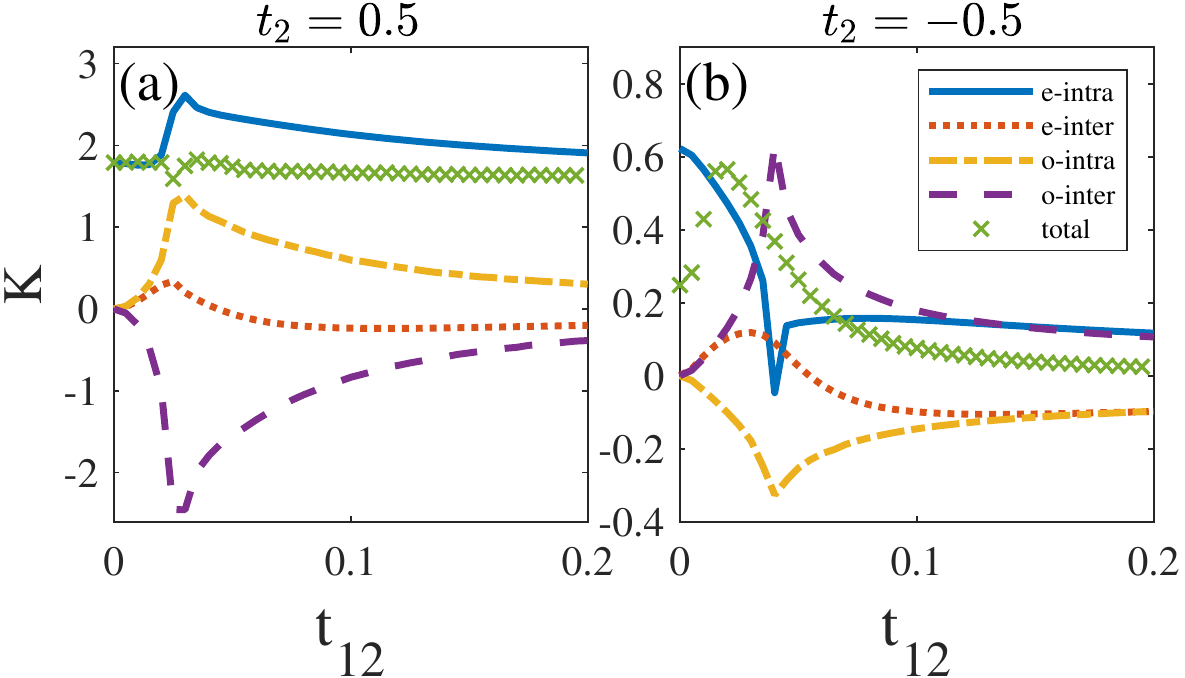}
\caption{Meissner effect as a function of orbital hybridization $t_{12}$ divided into even-frequency intra-band (blue solid), even-frequency inter-band (red dotted), odd-frequency intra-band (yellow dash-dotted), and odd-frequency inter-band (purple dashed) contributions, together with the total effect (green $\times$ marks) for (a) electron like band structure with $t_2 =0.5$, $\mu=0.3$ and (b) inverted band structure with $t_2=-0.5$, $\mu=0$. Here $\delta\mu=-0.2$, $\delta_+=0.03$, and $\delta_-=0.04$, creating a $s_{+-}$ state.
\label{fig:K-t12-t2pm}}
\end{figure}

In Fig.~\ref{fig:K-t12-t2pm} we plot the behavior of the even- and odd-frequency and intra- and inter-band contributions to the Meissner effect as a function of the $k$-independent orbital hybridization $t_{12}$. We further set $\delta_+=0.03$, $\delta_-=0.04$, $\delta\mu=-0.2$ and use in (a) an electron-like band structure with $t_{2}=0.5$ and $\mu=0.3$ and in (b) an inverted band structure with $t_{2}=-0.5$ and $\mu=0.0$ in order to comply with the parameter choices in the main text. These superconducting order parameters renders a $s_{+-}$ state with substantial odd-frequency pairing due to the large size of $\delta_-$.
The results show that for the region $t_{12} \lesssim 0.1$ all different contributions to the Meissner effects are reasonably large, while for larger $t_{12}$ the even-frequency intra-band contribution becomes dominant. This motivates our choice of $t_{12} \leq 0.1$ in the main text in order to capture the most interesting regime. Still, all conclusions drawn in the main text, in terms of both magnitudes and signs of the Meissner effect, hold true for the full range of $t_{12}$.

\begin{figure}[!thpb]
\centering
\includegraphics[width= 0.95\linewidth]{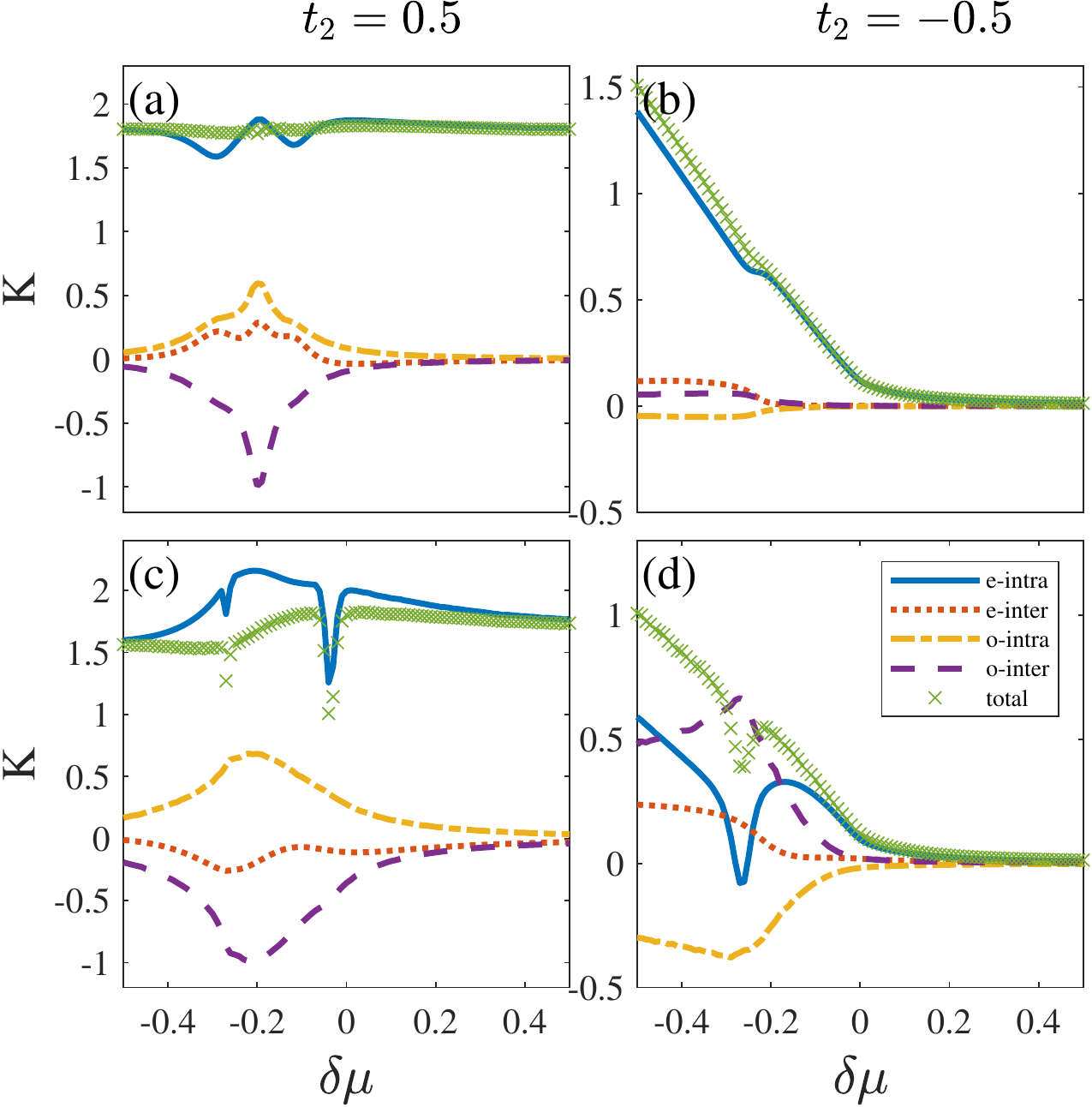}
\caption{Meissner effect as a function of orbital energy difference $\delta \mu$ divided into even-frequency intra-band (blue solid), even-frequency inter-band (red dotted), odd-frequency intra-band (yellow dash-dotted), and odd-frequency inter-band (purple dashed) contributions, together with the total effect (green $\times$ marks) for (a,c) electron like band structure with $t_2 =0.5$, $\mu=0.3$ and (b,d) inverted band structure with $t_2=-0.5$, $\mu=0$ and a momentum dependent orbital hybridization $\xi_{12}=t_{12}k^2$, with  (a,b) $t_{12}=0.05$, and (c,d) $t_{12}=0.2$. Here $\delta_+=0.03$ and $\delta_-=0.04$ creating a $s_{+-}$ state.
\label{fig:app1}}
\end{figure}
Next we investigate how the Meissner response changes if we instead use a momentum dependent orbital hybridization, $\xi_{12}(k)$.
In Fig.~\ref{fig:app1} we show the different contributions to the Meissner effect as a function of the orbital energy difference $\delta\mu$ for a $k$-dependent hybridization between orbitals of the form $\xi_{12}(k)=t_{12}k^2$. We further set $\delta_+=0.03$, $\delta_-=0.04$, creating a $s_{+-}$ state and report results for both an electron-like band structure (a,c) with $t_2=0.5$, $\mu=0.3$, and an inverted band structure (b,d) with $t_2=-0.5$, $\mu=0.0$. We also vary the hybridization strength, with $t_{12}=0.05$ (a,b) and $t_{12}=0.2$ (c,d). 
The results in these figures are in very good agreement with figures in the main text, Figs.~\ref{fig:K-dmu-t2-plus} and \ref{fig:K-dmu-t2-minus}, regarding the sign and magnitude of different contributions to the Meissner effect, and thus do not qualitatively change any of our conclusions. 
Exactly the same as in the main text, the odd-frequency intra- and inter-band contributions to the Meissner effect have opposite signs with respect to each other and also inverting the band structure flips the signs of these two contributions. Due to the opposite signs of the intra- and inter-band contributions, the total Meissner contribution from odd-frequency pairing is always small and hence even large odd-frequency pairing does never produce a destabilizing Meissner response, even for a momentum dependent orbital hybridization.
Additionally, we find for the case  $t_{12}=0.05$, that the odd-frequency contribution is relatively small, but it increases when we use $t_{12}=0.2$ in panels (c,d). This can easily be understood from the fact that the $t_{12}$-term is now multiplied by a $k^2$ dispersion, which reduces its effect at low momenta. Hence, in order to get a similar behavior to that of the $k$-independent case with $\xi_{12}=t_{12}$, we need to use larger values of pre-factor $t_{12}$.  
\\
\section{Band basis Hamiltonian}\label{app:bandbasis}
In this appendix, we transform the two-orbital Hamiltonian in Eq.~\eqref{eq:Ht} into the band basis where the kinetic energy is fully diagonal. While the original orbital basis of Eq.~\eqref{eq:Ht} is convenient for a lot of the derivations in the main text, the band basis gives a better intuition for some of our results. 

The normal-state Hamiltonian in the main text, $h(k)=\xi_+\tau_0+\xi_-\tau_z+\xi_{12}\tau_x$, is diagonalized and thus transformed into the band basis with the rotation matrix

\begin{align}
\hat{R}_{\theta/2}=\begin{pmatrix}
\cos(\theta/2) & -\sin(\theta/2)\\ \sin(\theta/2) & \cos(\theta/2)
\end{pmatrix},
\end{align}
where $\theta=\sin^{-1}\left(\xi_{12}/\sqrt{\xi_{12}^2+\xi_-}\right)$.
Applying this rotation matrix also to the superconducting part of Eq.~\eqref{eq:Ht} we retrieve the Hamiltonian in the band basis as:
\begin{align}\label{eq:Hb}
&\hat{H}_b=\nonumber\\ &\begin{pmatrix}
\epsilon_+&0&\delta_++\delta_-\cos(\theta)&
-2\delta_-\cos^2(\frac{\theta}{2})\\
0&\epsilon_-&-2\delta_-\sin^2(\frac{\theta}{2})&
\delta_+-\delta_-\cos(\theta)\\
\delta_++\delta_-\cos(\theta)&
-2\delta_-\cos^2(\frac{\theta}{2})&-\epsilon_+&0\\
-2\delta_-\sin^2(\frac{\theta}{2})&
\delta_+-\delta_-\cos(\theta)&0&-\epsilon_-
\end{pmatrix}
\end{align}
The Hamiltonian $H_b$ in the band basis nicely illustrates two features. One, the only coupling between the two bands is through the inter-band pairing terms, proportional to $\delta_-$. Thus, in the $\delta_- \rightarrow 0$ limit, the superconducting part of the Hamiltonian $H_b$ would commute with the normal part. As a result, odd-frequency pairing is a direct consequence of this inter-band term  \cite{Triola2020reviewAnnPhy}. This shows that odd-frequency pairing from the inter-band pairing is fundamentally the same, and only differ with a basis change, from the odd-frequency pairing generated by orbital hybridization $\xi_{12}$ and orbital pairing asymmetry $\delta_-$ used in the main text. Two, the intra-band pairing terms becomes explicit as $\delta_+ \pm \delta_-\cos(\theta)$ and it is easy to see when one of the bands become non-superconducting, a result used in the main text to explain some of the even-frequency intra-band results.

\bibliography{Ref.bib}

\end{document}